\documentstyle[aps, prb, preprint]{revtex}
\begin{document}
\draft
%
%


\title{Superconducting fluctuations and anomalous diamagnetism
in underdoped YBa$_2$Cu$_3$O$_{6+x}$ from magnetization and $^{63}$Cu 
NMR-NQR relaxation measurements}


\author{
P. Carretta,  A. Lascialfari,  A. Rigamonti,  A. Rosso and A. Varlamov 
\footnote{ On leave of absence from Institute of Steel and Alloys, 
 University of Moscow }} 


\address{
Department of Physics ``A. Volta'',  Unit\'a INFM di Pavia-
Via Bassi,  6  I-27100 Pavia,  ITALY}
\date{\today}
\maketitle


\widetext


\begin{abstract}






Magnetization and $^{63}$Cu NMR-NQR relaxation measurements are used to study
the superconducting fluctuations in YBa$_2$Cu$_3$O$_{6+x}$ (YBCO)
oriented
powders. In optimally doped YBCO the fluctuating 
negative magnetization $M_{fl}(H, T)$ is rather well described by 
an anisotropic Ginzburg-Landau (GL)
functional and the curves $M_{fl}/\sqrt{H}$
cross at $T_c$.
In underdoped YBCO, instead, over a wide temperature range an anomalous
diamagnetism is observed,
stronger than in the optimally doped compound by about an order of magnitude. 
The field and temperature dependences of $M_{fl}$ cannot
be described
either by an anisotropic GL functional or on the basis of scaling arguments.
The 
anomalous diamagnetism is more pronounced in samples with a defined order
in the Cu(1)O chains.
The $^{63}$Cu(2) relaxation rate shows little, if any, field dependence in the vicinity
of the transition temperature $T_c(H=0)$.  It is argued how the results in
the underdoped
compounds can be accounted for by the presence of charge inhomogeneities,
favoured by chains ordering.


\end{abstract}










\narrowtext








\section{Introduction}
The superconducting transition in conventional superconductors is rather
well described by mean field 
theories,  essentially because in the coherence volume $\xi^3$
a large number of pairs is present. 
On the contrary, in cuprate superconductors the small coherence length
$\xi$,  the reduced carrier
density,  the marked anisotropy and the high transition temperature $T_c$
strongly enhance the
superconducting fluctuations (SF). In a wide temperature range,
which can extend up
to $10$ or $15$ K above $T_c$, a variety of phenomena
related to SF 
\cite{bib1,bib2,bib3b} can be detected by several experiments such as
specific heat 
\cite{bib3}, 
thermal expansion \cite{bib4},  penetration depth \cite{bib5},  conductivity
\cite{bib6} and
magnetization \cite{bib7} measurements.
Also  NMR-NQR relaxation has been used  to detect 
 SF \cite{bib8,bib9,bib10,bib11,bib12}.
Recently the field dependence of the NMR relaxation rate $W$ and of the
Knight shift
have been studied, by varying the magnetic field from $2$ up to $24$ Tesla
\cite{bib10,bib11}.
Mitrovic et al. \cite{bib10} attributed the field dependence of $W(H)$ to the
corrections in the density of states (DOS) contribution to the spin-lattice relaxation. 
On the other hand,  Gorny et al. \cite{bib12}  have observed  $W(H)$ 
field independent,  up to 14 Tesla,  from 150 K to 90 K,  in a sample
showing a decrease in $1/T_1T$ starting
around $110$ K. 
An  analysis of the
role of SF in NMR experiments and of the field dependence of $W$ has
been recently carried out 
by Eschrig et al. \cite{bib15}, by extending the analitical approach of
Randeria and Varlamov \cite{bib16} to include short
wave-length and dynamical fluctuations.  






Of particular interest is the role of SF in underdoped high temperature
superconductors.
In fact,  in these compounds one has the peculiar phenomena of the opening,
at $T^*\gg T_c$,  
of a spin gap (as evidenced by NMR and neutron scattering)
and of a pseudo-gap
(as indicated by transport and ARPES experiments) \cite{bib17,bib18,bib19}. These
gaps,  which are
possibly connected \cite{bib19b},  have been variously related to SF of particular character  
\cite{bib20,bib21,bib22}.





In this paper 
magnetization and $^{63}$Cu(2) NMR-NQR relaxation measurements in underdoped YBCO are reported
and compared with the ones in the optimally doped compound. 
The magnetization data yield
information on the fluctuating diamagnetism (FD),  while the nuclear
relaxation rates $W$ convey
insights on the effects of SF on the ${\bf k}$-integrated spin susceptibility 
in
the zero frequency limit. 





The paper is organized as follows. In Sect. II some basic equations
describing  SF
and FD are recalled. After some experimental details (Sect. II), in
Sect. III the experimental findings and their analysis are presented, with 
emphasis on the effect of
the field on $W$ and on
the behavior of the diamagnetic magnetization in chain-ordered underdoped YBCO.
 The main results and conclusions are summarized in Sect. IV.





\section{Theoretical framework and basic equations}





  
Because of SF the number of Cooper pairs per unit volume,  which is
given by the average value of the square of the modulus of the order parameter 
$\sqrt{<\vert \psi\vert^2>}$, 
is different from zero above $T_c$. 
In the time-dependent Ginzburg-Landau (GL) description \cite{bib2,bib3b} the
collective amplitudes and the correspondent decay times of SF are given by
\begin{equation}
<\vert \psi_{\bf k}\vert^2>= {<\vert \psi_0\vert^2>\over 1+ \xi^2 k^2}
\hskip 12 pt {\rm and} \hskip 12 pt
\tau_{\bf k}={\tau_{GL}\over 1+ \xi^2 k^2}
\end{equation}
 where $\tau_{GL}=(\pi\hbar/8k_BT_c)\epsilon^{-1}$ and
$\xi(T)=\xi(0)\epsilon^{-1/2}$
, with $\epsilon= (T-T_c)/T_c$.
$<\vert \psi_{\bf k}\vert^2>$
plays the role of the Fourier components of
the average number of pairs per unit volume, while
$\tau_{\bf k}$ is the correspondent relaxation time.






Fluctuating pairs can give rise to a diamagnetic magnetization above $T_c$.
The diamagnetic magnetization $M_{fl}$ in general is not linear in
the field $H$. The curves $M_{fl}$ vs. $H$ and $M_{fl}$ vs. $T$ can be
tentatively analized by generalizing the
 Lawrence-Doniach model,
  using the free energy functional \cite{bib23,bib24}
\begin{eqnarray}
F[\psi]=\sum_n\int d{\bf r}\biggl[ a\vert \psi_n\vert^2 + {b\over 2}\vert
\psi_n\vert^4 +  \nonumber \\
+{\hbar^2\over 
4 m_{\parallel}}\vert[{\bf \nabla_{\parallel}}- {2ie\over\hbar c}{\bf
A_{\parallel}}] \psi_n\vert^2 +
 t \vert \psi_{n+1}- \psi_n\vert^2 \biggr]
\end{eqnarray}
where the last term takes into account the tunneling coupling between
adjacent layers. 
From Eq. 2, deriving the free energy in
the presence of the field and by means of a numerical derivation
($M_{fl}=-\partial F/\partial H$), one can obtain the fluctuating magnetization.




According to scaling arguments \cite{bib25,bib26,bib27}
for moderate anisotropy
  (quasi-3D case) one expects that the magnetization curves,  at constant
   field,  cross at $T_c(H=0)$ when  
 the magnetization is scaled by $\sqrt{H}$. 
 The amplitude ${M_{fl}}$ at $T=T_c$ departs 
 from the Prange's result \cite{bib28} by a factor $3-7$,  corresponding to
  the anisotropy ratio $(\xi_{\parallel}/\xi_{\perp})$. 
 Accordingly \cite{bib29},  the data at 
 different fields collapse onto an universal curve
  when $M_{fl}/T$ is reported as a function of $\epsilon/H^{1\over{2\nu}}$, 
for a critical exponent for the coherence length corresponding to $\nu\simeq 0.66$.


For strong anisotropies,  namely quasi-2D systems,   the $M_{fl}$
curves at constant field cross 
each other at $T_c(0)$. ${M_{fl}}(T_c)$ is larger than the value obtained
 from the Gaussian approximation by a factor  \cite{bib27} around $2$.
Collapse of 
 the data onto a universal curve occurs for $\nu$ well above $0.66$,  usually in the
range $1.2- 1.4$.





The contributions to the relaxation rates W due to SF can be derived 
within a Fermi liquid scenario,  
withouth specifying the nature of the interactions \cite{bib2}.
The direct and most singular contribution,  equivalent to the
Aslamazov-Larkin paraconductivity, 
 is not effective as nuclear relaxation mechanism. 
 The positive Maki-Thompson (MT) contribution $W_{MT}$
results from a purely quantum process, 
  involving  pairing of the electron with itself at
   a previous stage of motion,  along intersecting trajectories.
A negative SF contribution  $W_{DOS}$ comes from the density of states 
 reduction when electrons are subtracted to create pairs. Approximate
expressions 
 for $W_{DOS}$ and $W_{MT}$  
 can be derived by resorting to 
simple physical arguments as follows. The relaxation rates can be approximated in the form
\begin{eqnarray}
 2W\simeq {\gamma^2\over 2}\langle A_{\vec k}\rangle k_BT\sum_{{\vec
k}}\left(\frac{{\chi ^{\prime \prime }}_{spin}(
{\vec k},\omega)_{\perp}}{\omega}\right)_{\omega \rightarrow 0}\simeq
\frac{\gamma^2}{2N} \langle A_{\vec k}\rangle
 {k_B}T\chi_{spin} (0,0)\sum_{\vec k} \tau_{\vec k}
\end{eqnarray}
where  $ <A_{\vec k}>$     is an average form factor for Cu nuclei and
$\tau_{\vec k}   = J_{\vec k}(0)$   an effective spectral
density \cite{bib17,bib30}.
SF modify the static spin 
  susceptibility $\chi_{spin}(0,0)$ and the effective correlation time  $\sum_{\vec k}
\tau_{\vec k}$ in Eq. 3.   
From Eq.1 the in-plane  density of pairs  is 
$ n_{c}= \sum_{\vec k}
\langle |\psi_{\vec k}|^2\rangle=2 {n}_e({k_B T_c}/ E_F) \ln{1 \over \epsilon}$
and therefore the SF imply


\begin{eqnarray}
\chi_{spin}(0,0)=
{n_e\mu_B^2\over E_F}
\left[ 1-\frac{2k_B T_c}{E_F}\ln \frac{1}{\epsilon}\right]
\end{eqnarray}
For the dynamical part in Eq. 3,  the DOS contribution 
	can be  obtained by averaging over the
	 BZ the collective correlation time in  Eq. 1:
\begin{eqnarray}
\langle \tau _{DOS}\rangle
 =\sum_{{\vec k}} \tau_{\vec k}^{DOS}
= \hbar \rho (E_F)\frac{\hbar}
{4 E_F \tau}
\ln \frac{1}{\epsilon}  
\end{eqnarray}
in the dirty limit,  differing from the exact calculation  \cite{bib16}
only by a small
 numerical factor. $\tau$ is the electron collision time.
  The MT contribution  has to be evaluated in the framework of diagrammatic
theories \cite{bib2}.
  The final result corresponds to the 2D-average over the BZ of a decay rate of
diffusive character 
 $\Gamma_{\vec k} = D k^2$     
 ( with  $D=  {v _F} ^2  {\tau}/{2}$    carrier diffusion constant),
 phenomenologically  accounting for phase breaking processes by adding in
the decay 
 rate a frequency   ${\tau_{\phi}}^{-1}$:


\begin{eqnarray}
\langle \tau _{MT}\rangle =
\sum_{{\vec k}}
\{ (Dk^2+\tau _\phi ^{-1}) \epsilon ( 1 +
\xi^2 k^2)\}^{-1} = \nonumber\\ 
= \hbar \rho (E_F)\frac{\hbar}
{4 E_F \tau} \frac{1}{(\epsilon-\gamma_{\phi})}
\ln \frac{\epsilon}{\gamma_{\phi}}
\end{eqnarray}
where  $\gamma_{\phi} =  {\xi^2 (0)}/
{\tau_{\phi} D} = {\pi \hbar }/{8 K_B T_c \tau_{\phi}}$
 is a dimensionless pair breaking parameter.
 
 
By indicating with $W^0$ the relaxation rate in the
absence of SF  
and by neglecting the  correction to   $\chi_{spin}(0, 0 )$ 
 (Eq. 4), from Eqs. 3, 5 and 6 the relaxation rate in 2D systems turns out 
 

 
\begin{eqnarray}
W^{SF} = W^0
\left[  \frac{\pi \hbar }{8E_F\tau} \frac{1}{(\epsilon-\gamma_{\phi})}
\ln \frac {\epsilon}{\gamma_{\phi}}-\frac{0.8\hbar}{E_{F} \tau}
\ln\frac{1}{\epsilon}\right]  \label{ram}
\end{eqnarray}



To extend this equation to a layered system,  when dimensionality 
crossover (2D$\rightarrow$ 3D) occurs,  one has to substitute \cite{bib2}
$  ln({1 \over \epsilon})$   by  $2ln\left[2/(\sqrt {\epsilon}+\sqrt{
 \epsilon + r})\right]$ and $ln({\epsilon  \over \gamma_{\phi}})$   by
    $2ln\left[\frac{(\sqrt {\epsilon}+\sqrt{ \epsilon + r})}
    {(\sqrt {\gamma_\phi}+\sqrt{ \gamma_{\phi}+ r})}\right]  $,
 with  $r= 2  {\xi_c}^2 (0) /d^2$ anisotropy parameter ($d$ is the interlayer distance).
  It is noted that the MT contribution   (first term in Eq. 7)
   is present only for s-wave orbital pairing,  while it is averaged to
almost zero for d-symmetry.
   



\section{Experimental Details}


The measurements have
 been carried out in  oriented powders of  optimally doped YBCO in one chain-disordered 
underdoped and in two chain-ordered underdoped YBCO samples \cite{bib31}
  The oxygen content in the underdoped compounds was close to 6.66,  
with sligth differences  in $T_c$. Electron diffraction microscopy evidenced 
  the expected tripling of the a-axis,  while resistivity measurements show a 
 sharp transition with zero resistivity at T= 62 K and 
occurrence of paraconductivity below about 75 K \cite{bib31}.
 Table I collects the main properties of the samples, 
  as obtained from a combination of measurements.














The $^{63}$Cu relaxation rates $2W\equiv {T_1}^{-1} $
 have been measured by standard pulse techniques.
In NQR the recovery towards the
 equilibrium after  the saturation of the $(\pm \frac{1}{2}
  \rightarrow \pm \frac{3}{2})$
  line is  well described by an exponential law, directly yielding $6W$.
  For the NMR  relaxation  a good alignment of the $\vec c$ axis of the grains 
  along the magnetic field is crucial to extract
   reliable values of W (H). The NMR satellite line, 
    corresponding to the  $(\frac{1}{2} \rightarrow \frac{3}{2})$
   transition can be used to adjust the alignment
    and to monitor the spread in the orientation of the c-axis,  
    the resonance frequency 
   being shifted at the first order by the term  $eQV_{zz} ( 3 cos^{2}\theta
-1 )/ h$, 
    due to the quadrupole interaction ( $\theta $  angle of the $\vec
c$-axis with the field).
     From the  width  of the spectrum (Fig. 1a)
      the spread in the orientation of the c-axis appears within about two
degrees.



The recovery law for the NMR satellite transition is
\begin{equation}
   y(t)=0.5 exp(-12 Wt) + 0.4 exp(-6Wt) + 0.1 exp(-2Wt)     
\end{equation}
This law has been checked to fit well the experimental data (Fig. 1b), proving the lack 
    of  background contamination due to  other resonance lines.
     



The magnetization has been measured by means of a Metronique Ingegnerie SQUID
magnetometer, 
on decreasing the temperature at constant field and, at selected
temperatures,  by varying $H$. 
The paramagnetic contribution  to M was obtained from  M
vs $H$ at T $\geq$ 110
 K,  where practically no  fluctuating magnetization is present.
Then $M_{fl}$ was derived
  by subtraction, singling out a
 small contribution from paramagnetic impurities.
The paramagnetic susceptibility turns out little temperature dependent around $T_c$
and this dependence will be neglected in discussing the much stronger diamagnetic term. 


In optimally doped YBCO the measurements have been carried out  at the purpose to confirm  
the results  recently obtained in single crystals
 by other authors\cite{bib23,bib27,bib29,bib36}. 
 The isothermal magnetization curves are satisfactorily described on the basis of the
anisotropic GL functional ( Eq. 2), for
$\epsilon \ge 4\cdot 10^{-2}$.
Only close to $T_c$ and for $ H\le  1$
Tesla an 
observable 
 departure is detected,  indicating  crossover to  a region of non-Gaussian SF, in agreement with 
recent thermal expansion measurements \cite{bib4}.
          The $M_{fl}$ vs. $T$ curves  cross at $T_c(0)$ when
$M_{fl}$ is scaled by
      $\sqrt H$, as expected  for moderate
anisotropy \cite{bib27}.





\section{Results and Analysis}


Let us first comment  the data
in optimally doped YBCO (Fig. 2a). The comparison between the NMR and NQR $W$'s has 
been already discussed in previous
papers \cite{bib8,bib9}. 
Here we  only add a few comments   
motivated by more recents works  \cite{bib10,bib12,bib39} 
involving the remarkable aspect 
of the field dependence.


The NQR $W$ 
can be compared to Eq. 7, 
 by using the values \cite{bib2}   $\tau= 10^{-14} s$, 
 $\tau_{\phi} = 2.10^{-13} s$ 
and a dephasing time parameter $\gamma_{\phi} = 0.2$. A quantitative fitting
is inhibited by the
fact that the background contribution to W does not follow the Korringa law,
in view
of correlation effects among carriers \cite{bib17}. A firm deduction, however, is that
an anisotropy parameter  $r\neq 0$,  
and namely a crossover to a 3D regime,  is required to avoid 
unrealistic values (2D line in Figure 2a), as indicated also by
magnetization measurements (see later on).



 Mitrovic et al. \cite{bib10} have discussed the field dependence of
$T_1$ in terms of  
quenching of the DOS term only,  by resorting to the theory of Eschrig et
al. \cite{bib15}.
The field dependence of the SF contribution to the  relaxation rate
is a delicate 
issue, 
because of the nontrivial interplay of many parameters which include reduced
temperature $\epsilon$, 
reduced field $\beta = \frac{2H}{H_{c2}} $,  anisotropy parameter $r$,
elastic and phase breaking times
$\tau$ and $\tau_{\phi}$.
Eschrig et al. \cite{bib15} have extended the analitical approach
\cite{bib16} by taking into account
arbitrary values of $(K_B T \tau / \hbar)$,  short-wave fluctuations and
dynamical fluctuations.
The price of this  generalization is the restriction to a 2D spectrum
of SF 
fluctuations.
In view of our experimental findings  the analysis of the field effect in a
purely 2D framework is
questionable.
On the other hand,  a re-examination
 of the field effect for a layered system is now possible by
applying the method of the DOS term regularization devised  by Budzin and
Varlamov \cite{bib39},
which has indicated how the divergences can be treated. 
Since the dynamical fluctuations could be relevant
only for fields of the order of
$H_{c2} $,  up to $20 - 30 $ $Tesla$ the fluctuations can be safely treated
in the nearly static limit.
The field dependence for weak fields ($\beta\ll\epsilon$) turns out \cite{bib40}



\begin{eqnarray}
W(\beta\ll\epsilon)-W(0,\epsilon) 
= W^{(0)}\frac{\hbar}{24 E_F\tau}\biggl(\kappa(T\tau)-{\pi\over 8\gamma_{\phi}} \biggr)
 \frac{\epsilon+r/2}{[\epsilon(\epsilon+r)]^{3/2}}\beta^2
\end{eqnarray}


where


$\kappa (T\tau )=\frac{7\zeta (3)}\pi \frac 1{4\pi T\tau \left[ \psi
(1/2)-\psi (1/2+1/4\pi T\tau )\right] +\psi ^{\prime }(1/2)}= 
\Bigg\{
\begin{tabular}{cc}
 $14\zeta (3)/\pi ^3$, & $T\tau \ll 1$ \\ 
 $4T\tau$ ,  & $1\ll T\tau \ll 1/\sqrt{\epsilon}$%
\end{tabular}
$
\vskip 16pt
(here $\hbar= k_B=1$).


 So both effects of  increase and of decrease of $W$ on increasing the 
magnetic field are possible, depending on the mean 
free path in the specific sample. If $T\tau\lesssim 0.1$ the main correction
to $W$ is due to the MT term and one should observe $W$ decreasing with 
increasing field, while if $T\tau\gtrsim 0.1$ the DOS correction becomes
dominant and $W$ is expected to increase.


As it appears from Fig. 2 the effect of the field for $T\geq T_c(H=0)$ is small, if
any. It is noted that if Eq. 7 is applied to the underdoped compounds the reduction
in the Fermi energy $E_F$ and the increase in the anisotropy (i.e. decrease of $r$) would imply a 
sizeable increase in $W^{SF}$. This  enhancement  in the underdoped compounds does 
not  occur (see Fig. 2b and 2c for the chain-disordered and chain-ordered YBCO's, 
respectively), unless the decrease of $W$ over a wide temperature range, usually related to 
the spin-gap opening \cite{bib17}, should be attributed to a field-independent DOS term.
On the other hand, the conventional SF of GL character should occur only close to $T_c$, where,
on the contrary, $\chi_{spin}(0,0)$ is  little temperature dependent.
Eq. 9, in principle, does predict a field dependence. However, if typical values $E_F\simeq 3000$ K,
$\tau\simeq 10^{-14}$, $(\kappa(T\tau)-(\pi/ 8\gamma_{\phi})\simeq 1$ and $r\simeq 0.1$ are used,
for
 $\epsilon\simeq 3\times 10^{-2}$ one has $W(\beta)-W(0)\simeq 0.15 W^{(0)}\beta^2$, hardly to evidence
for $\beta\ll 1$. A comprehensive discussion of the field dependence of the SF contribution 
to the  relaxation rate is given elsewhere \cite{bib40}. Here we only remark  that the
results in strong fields ($\beta\gtrsim 0.2$) from Mitrovic et al. \cite{bib10} can hardly be
justified on the basis of Eq. 9.
 


As regards the magnetization  measurements in optimally doped YBCO, as already mentioned,
our data 
indicate a 3D regime crossing from Gaussian to critical fluctuations close to $T_c$. The value
of $\gamma_{an}
 m_3  (\infty)$  
\cite{bib27} at the crossing point of the curves $M_{fl} / \sqrt{H} $ vs. $T$ 
turns out around $-1.5$, 
corresponding to an anisotropy factor   $\gamma_{an} =4.5$ . A collapse onto a
common function is 
obtained when $M_{fl}/ {\sqrt{H} T}$ is plotted as a function of     
$\epsilon/ H^{0.747}$.
The collapse   
fails in magnetic fields less than about 0.3 Tesla. A small-field departure
from the 
universal function has been already observed in underdoped compounds
\cite{bib29}, 
but no mention about it is found in the literature for optimally doped YBCO.



In the underdoped compounds 
 a marked enhancement of the fluctuating magnetization is observed (Figs. 3 and 4).
In Fig. 4
  the susceptibilty,  defined as $\frac{M}{H}$ for $H=0.02$ Tesla, is reported.
   The inset is a blow up of the results around the temperature  where 
    reversing of the sign of the magnetization occurs. The susceptibilities for
chain-ordered and chain-disordered underdoped  samples
    are compared with the one in optimally doped YBCO in Fig. 5.




In underdoped YBCO the magnetization curves dramatically depart from the 
ones expected on the basis of  Eq. 2 and numerical derivative.
No crossing point is observed in the 
curves $M_{fl}$ vs $T$ and no collapse is found on  anisotropic 3D or
 2D-like curves for ${M_{fl}}/{\sqrt{H}}$  or  $M_{fl}$ vs   ${\epsilon}/{H^{
1/2 \nu}}$ (Fig. 6).
 These anomalies are more manifested in the chain ordered compound,  as it 
  can be realized from the peculiar shape of the isothermal magnetization vs. $H$ (Fig. 7).
One could remark that a fraction of a few percent of non-percolating local
superconducting regions can account for the absolute value of the
diamagnetic susceptibility and for the shape of the magnetization curves.
A trivial
chemical inhomogeneity could be suspected. The fact that the anomalous diamagnetism
has been observed in three samples  differently prepared and since it is
strongly enhanced in the chain-ordered ones, supports the intrinsic
origin of the phenomenon.
Recent theories \cite{bib36b,bib37b} have discussed how a distribution of
local superconducting temperatures, related to charge inhomogeneities, can cause
anomalous diamagnetic effects.


Mesoscopic charge inhomogeneities have been predicted on the basis of
various theoretical approaches \cite{bib21,bib48}. In 
underdoped compounds one could expect the
occurrence of preformed Cooper pairs causing "local"
superconductivity lacking of long-range phase coherence. An inhomogeneous
distribution of carriers, on a mesoscopic scale, is supported by
$ab$ plane optical conductivity
\cite{bib45} and it is the basic ingredient of the stripes model \cite{bib49}.
 Since the anomalous diamagnetism is enhanced in chain-ordered underdoped 
YBCO one is tempted to relate it to the presence of stripes.
The insurgence of phase
coherence among adjacent charge-rich regions can qualitatively be expected
to yield strong screening above the bulk superconducting temperature 
\cite{bib36b,bib37b,bib41b}.
Therefore, one could phenomenologically describe the role of SF 
in underdoped compounds, as follows.
At $T\gg T_c$, just below $T^*$, fluctuating  pairs are
formed, without long-range phase coherence. Below $T_{SF}$, about
$20-30$ K above $T_c$, phase coherence among adjacent charge-rich drops \cite{bib37b}
start to develop, yielding the formation of superconducting
loops with strong
screening and causing the onset of the anomalous diamagnetism. Close to $T_c$, SF 
of GL character occur and long-range phase coherence sets in. 
 
 
\section{    Summarizing Remarks and Conclusions}



By combining magnetization and NMR-NQR relaxation measurements,  
 an attempt has been done to clarify the role of superconducting
fluctuations and 
  of fluctuating diamagnetism in underdoped YBCO,
 vis-a-vis the optimally doped compound. 
 In the latter case the 
 fluctuations above $T_c$  are rather well described
 by an anisotropic Ginzburg-Landau (GL) functional and by scaling 
    arguments for slightly anisotropic systems.
 A breakdown of the
 Gaussian approximation for small magnetic fields has been observed close to $T_c$. 
The $^{63}$Cu  
relaxation rates $W$ 
 around $T_c$ show  little field dependence, if any. One cannot rule out the 
presence of a MT contribution, and then of a small 
$s-$wave component in the
spectrum 
of the fluctuations, which should be sample-dependent in
view of the role of impurities in the pair-breaking mechanisms. 
 In underdoped  YBCO an anomalous diamagnetism 
       is observed,  on a large temperature range. The diamagnetic 
       susceptibility at $T_c$ is about an order of magnitude larger 
       than the one in the optimally doped sample and  the isothermal
        magnetization curves cannot be described by the anisotropic GL
functional. 
	Scaling arguments,  such as the search of an universal function in terms 
	of ${\epsilon}/{H^{1/2\nu}}$, appear inadequate to justify the experimental
findings.
	 The anomalies are more marked in the chain-ordered samples. Also in
the underdoped YBCO $^{63}$Cu $W$ in  NQR almost  coincide with the
NMR ones at $H=9.4$ Tesla, in agreement with a theoretical estimate
of the effect of the magnetic field.  Another conclusion
drawn from the field and temperature dependences of the $^{63}$Cu relaxation rate is that
the spin gap does not depend on the magnetic field and that the
behavior of the spin susceptibility for $T\gg T_c$
cannot be ascribed to SF of GL character. 


The  fluctuating diamagnetism observed in underdoped YBCO can be
phenomenologically
   accounted for by the presence of charge inhomogeneities at
    mesoscopic level.  
The  
anomalies in the magnetization curves can be attributed to "locally
superconducting" 
     non-percolating drops,  present above the bulk $T_c$ and favoured by chain
ordering. 
     



\section{Acknowledgements}



A. Bianconi, F. Borsa, J. R. Cooper and M. H. Julien  are thanked for useful discussions. 
The collaboration of Paola Mosconi in processing the SQUID
magnetization data is gratefully acknowledged.
Thanks are due to B.J. Suh and  P. Manca,  for having provided well
characterized
underdoped YBCO  compounds.  The  availability of  
the  Metronique Ingegnerie SQUID magnetometer from the Departement  of
Chemistry,
University of Florence (Prof. D. Gatteschi) is gratefully acknowledged.
 The research has  been carried out 
 in the framework of a "Progetto di ricerca avanzata (PRA)", sponsored by   
Istituto Nazionale di Fisica della Materia (PRA-SPIS 1998-2000).






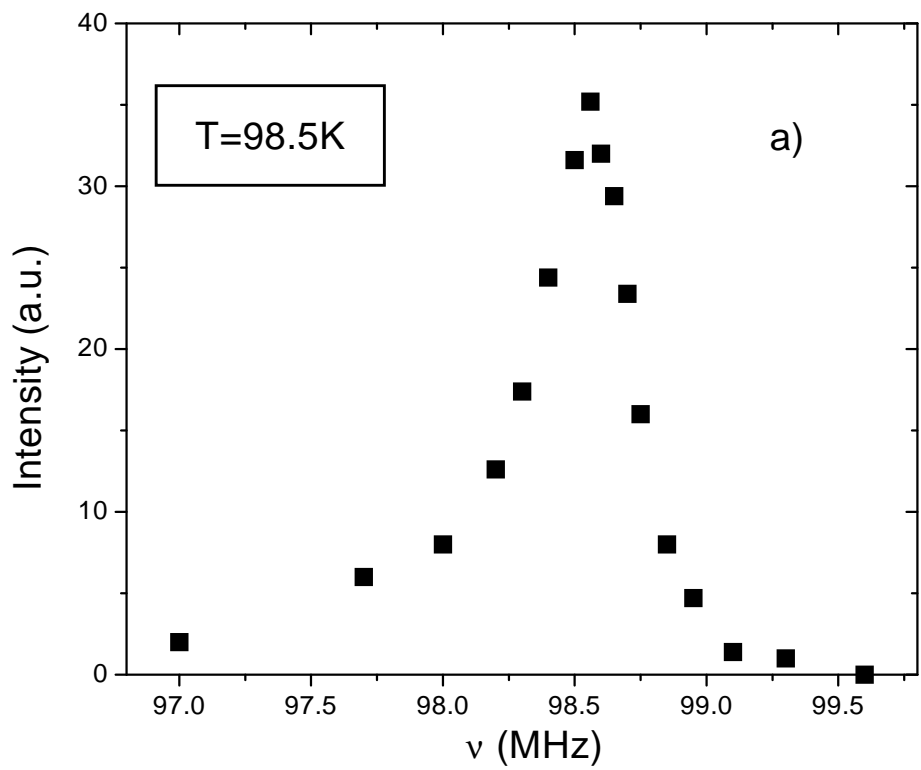
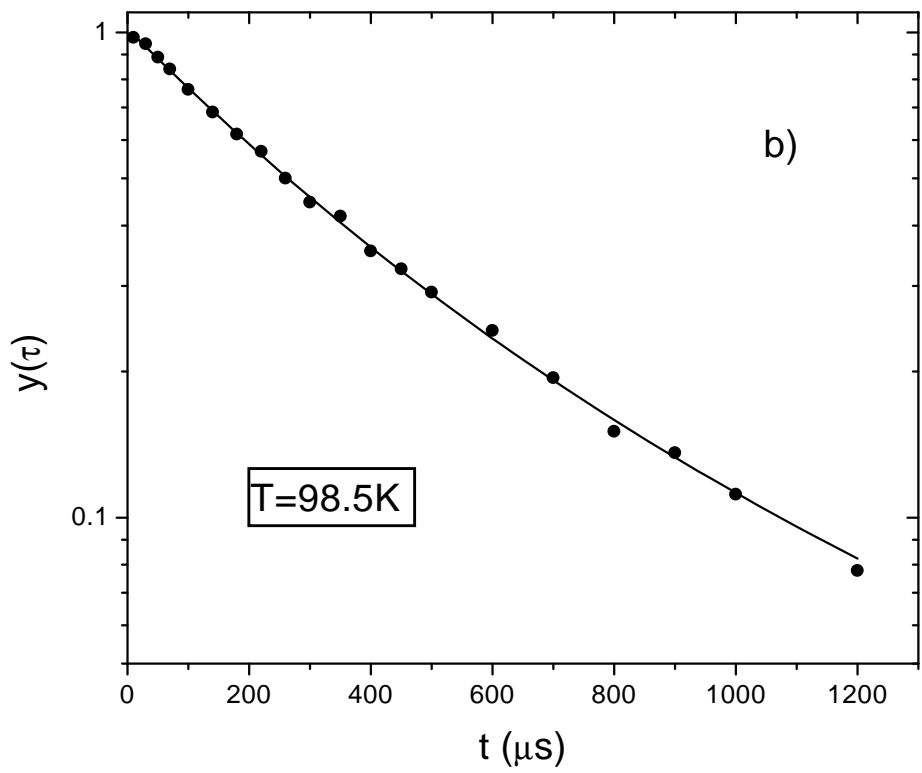
\begin{figure}
\caption {{\it a)} NMR spectrum of the $^{63}$Cu(2) satellite line and the correspondent
relaxation law ({\it b}) (the solid line is the best fit according to Eq. 8 in the text).}
\end{figure}




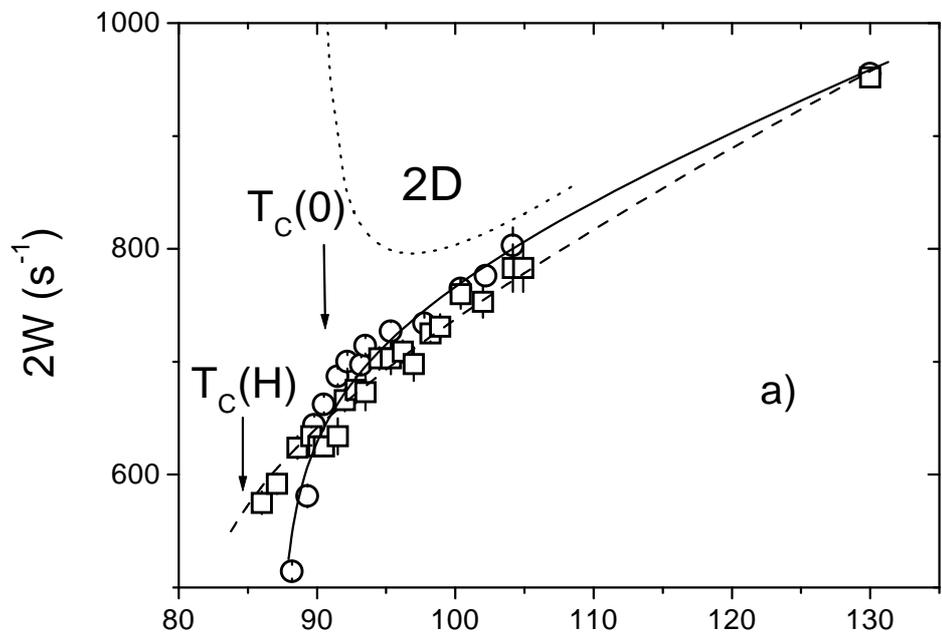
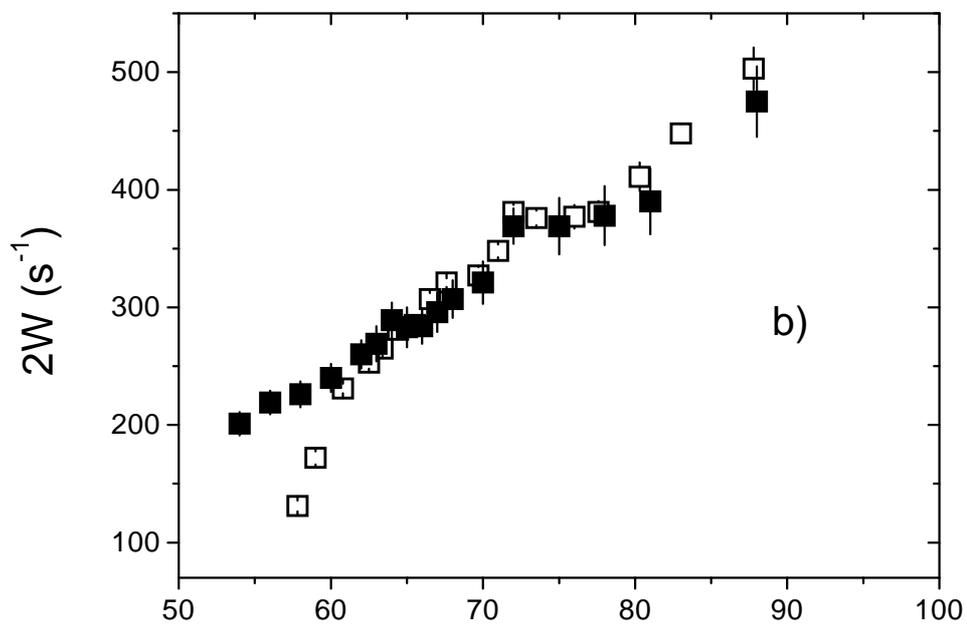
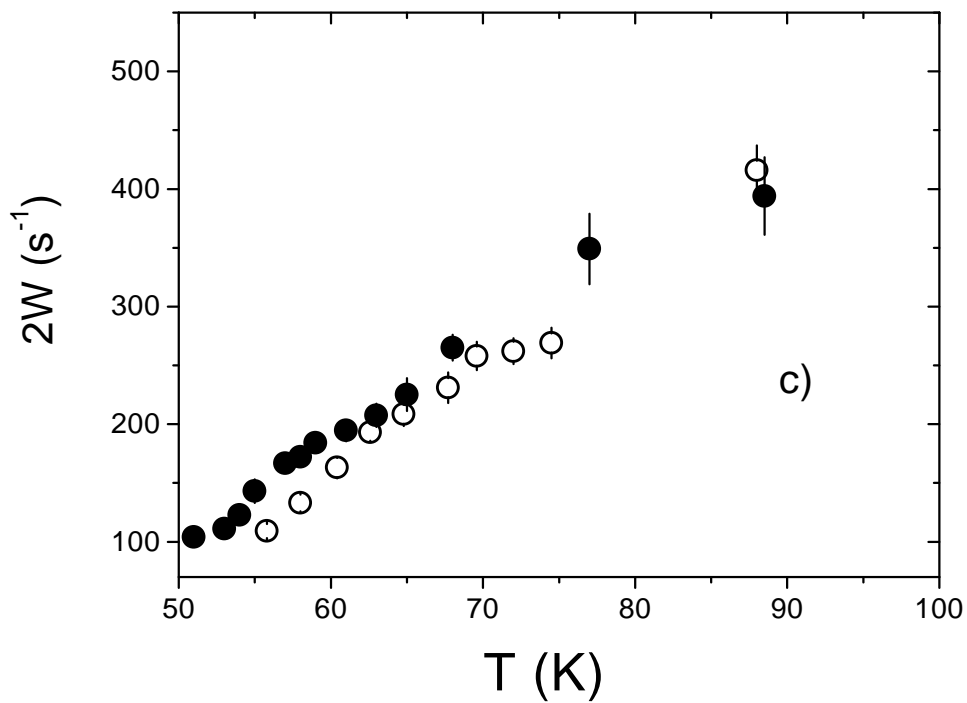
\begin{figure}
\caption {{\it a)} NQR (o) and NMR ($\Box$) ($H_o= 6$ Tesla $\parallel \vec c$) $^{63}Cu$ 
 relaxation rates  $2W$  in optimally doped YBCO. The solid and dashed lines are guides for the 
eye. The dotted line is the temperature behavior according to Eq. 7 in the text for $r=0$
(namely 2D spectrum of fluctuations) and $\gamma_{\phi}=0.2$.  NQR (empty simbols) and
NMR (full symbols) relaxation rates $2W$  (in a field of 9.4 Tesla along the c-axis) 
 in chain-disordered ({\it b}) and chain-ordered ({\it c}) underdoped YBCO around $T_c$.}
\end{figure}


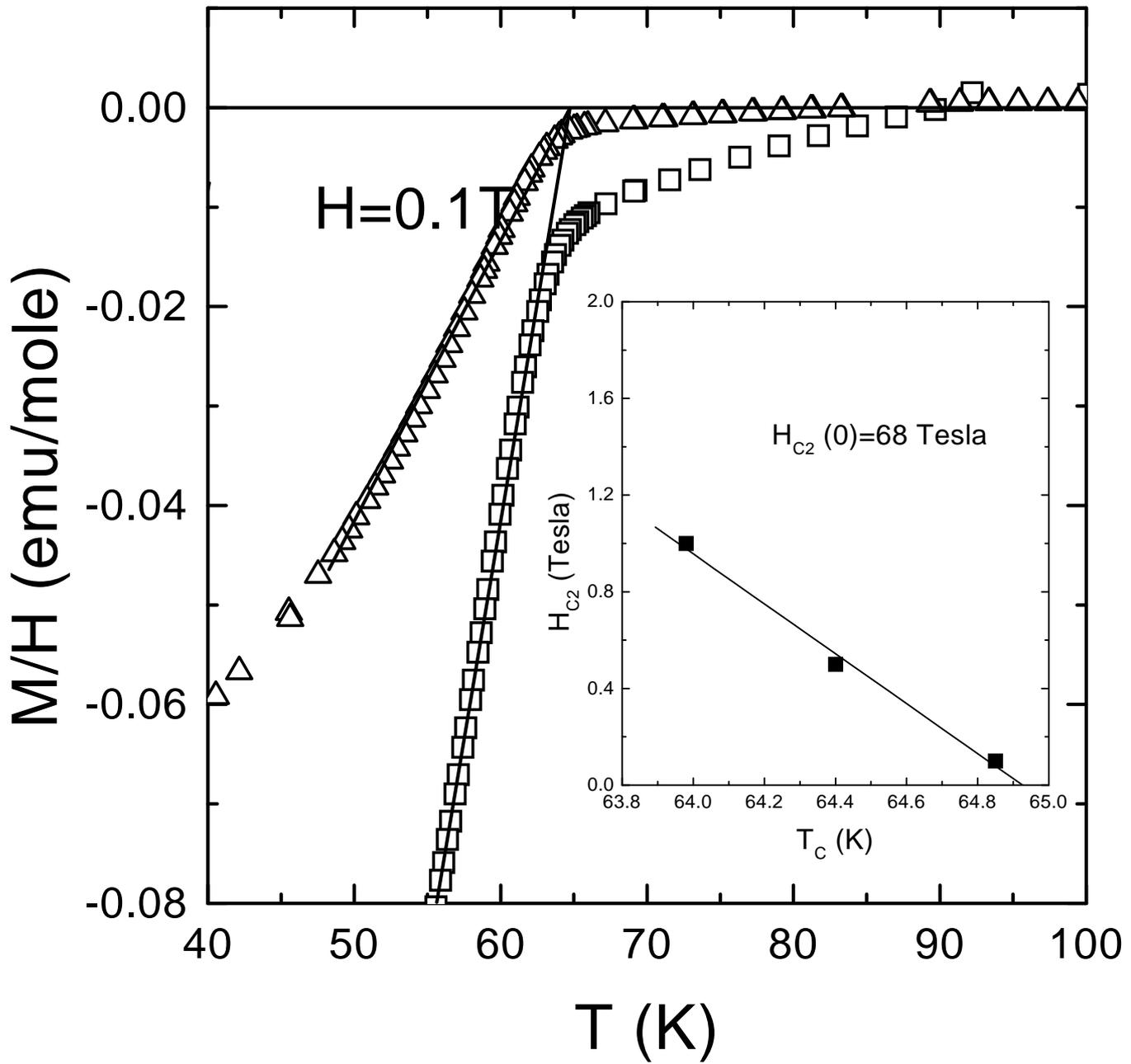
\begin{figure}
\caption { Raw data for $M/H$ ($H=0.1$ Tesla) in chain-ordered YBCO, 
for field along the c-axis   ( $\Box$ ) and in the a-b
plane
  ( $\triangle$ ). In the inset the results for $H_{c2}$ around $T_c$ are
shown (estimated
experimental error $\leq 2$ \%). The solid lines indicate the procedure to
evaluate $T_c(H)$ and $T_c(0)$.}
\end{figure}
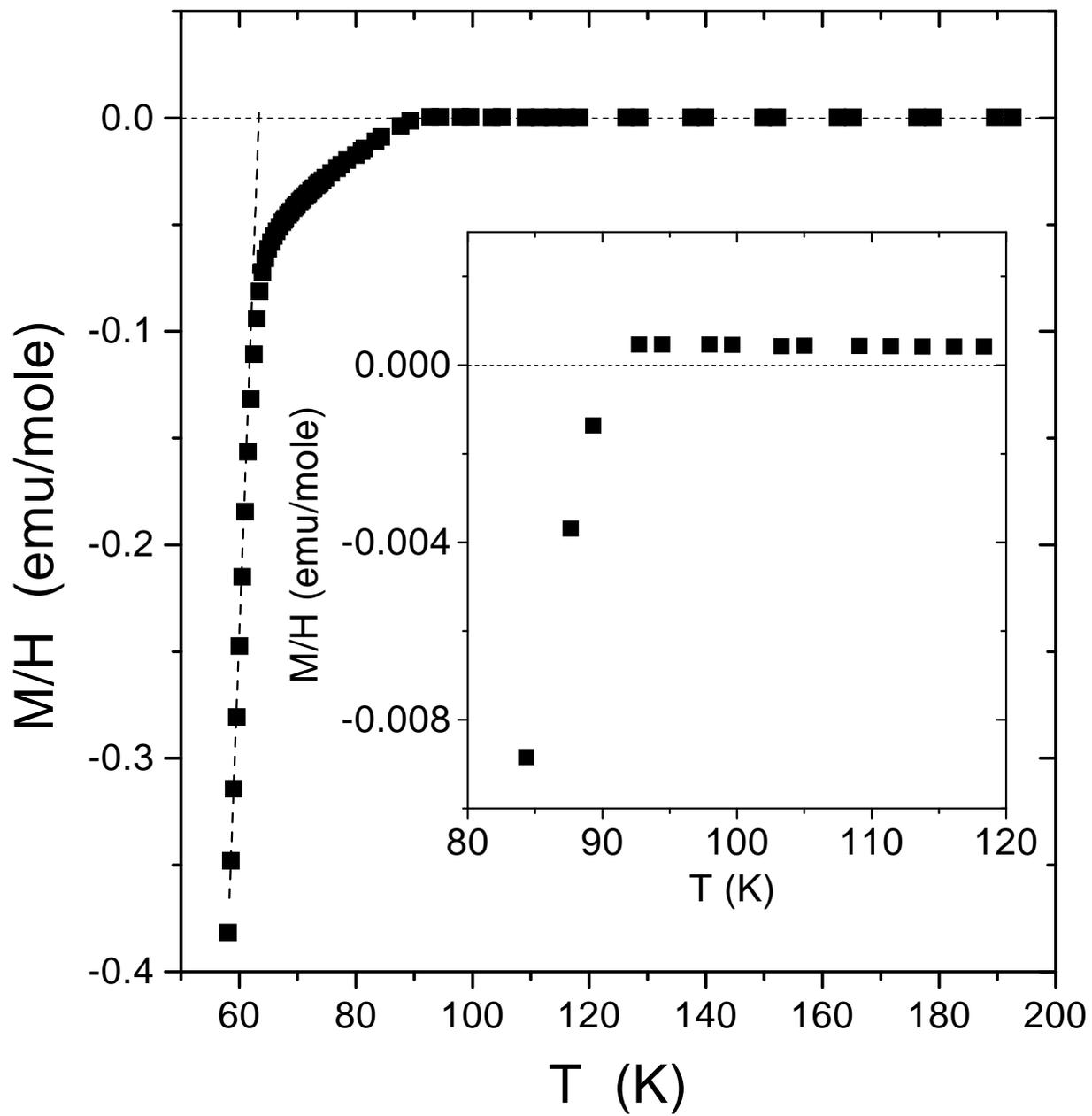
\begin{figure}
\caption { Ratio $M/H$, in a field of 0.02 Tesla along the c-axis, in chain-ordered YBCO.
The inset is the blow-up of the data for $T \ge 85 K$.
Similar results have been obtained in another sample prepared with the same procedure.}
\end{figure}
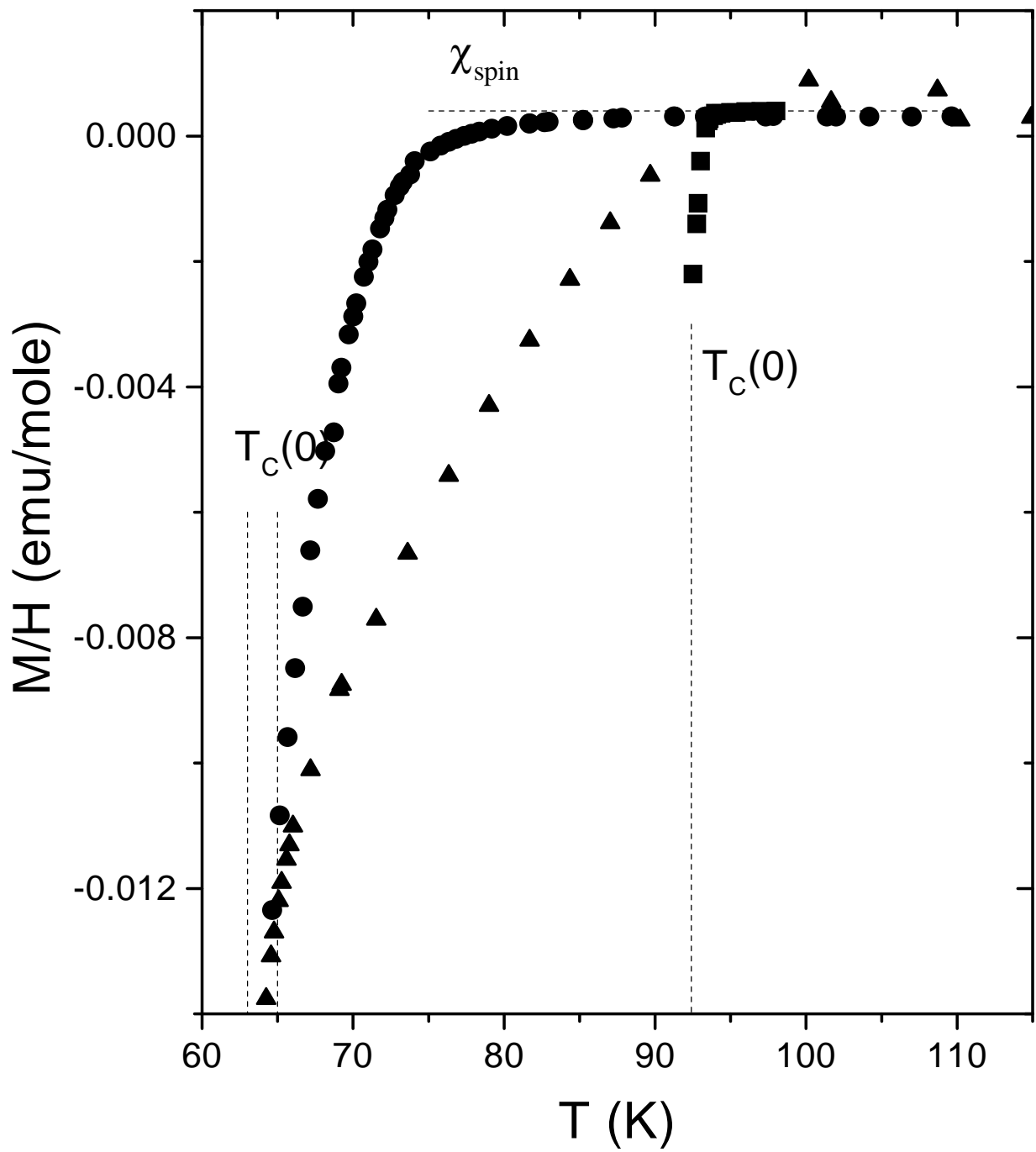
\begin{figure}
\caption { Comparison of the susceptibility (defined as $M/H$, for $H=0.02$ Tesla
along the c-axis)
in optimally doped YBCO (squares) and in chain-disordered (circles) and 
chain-ordered (triangles) underdoped YBCO.}
\end{figure}
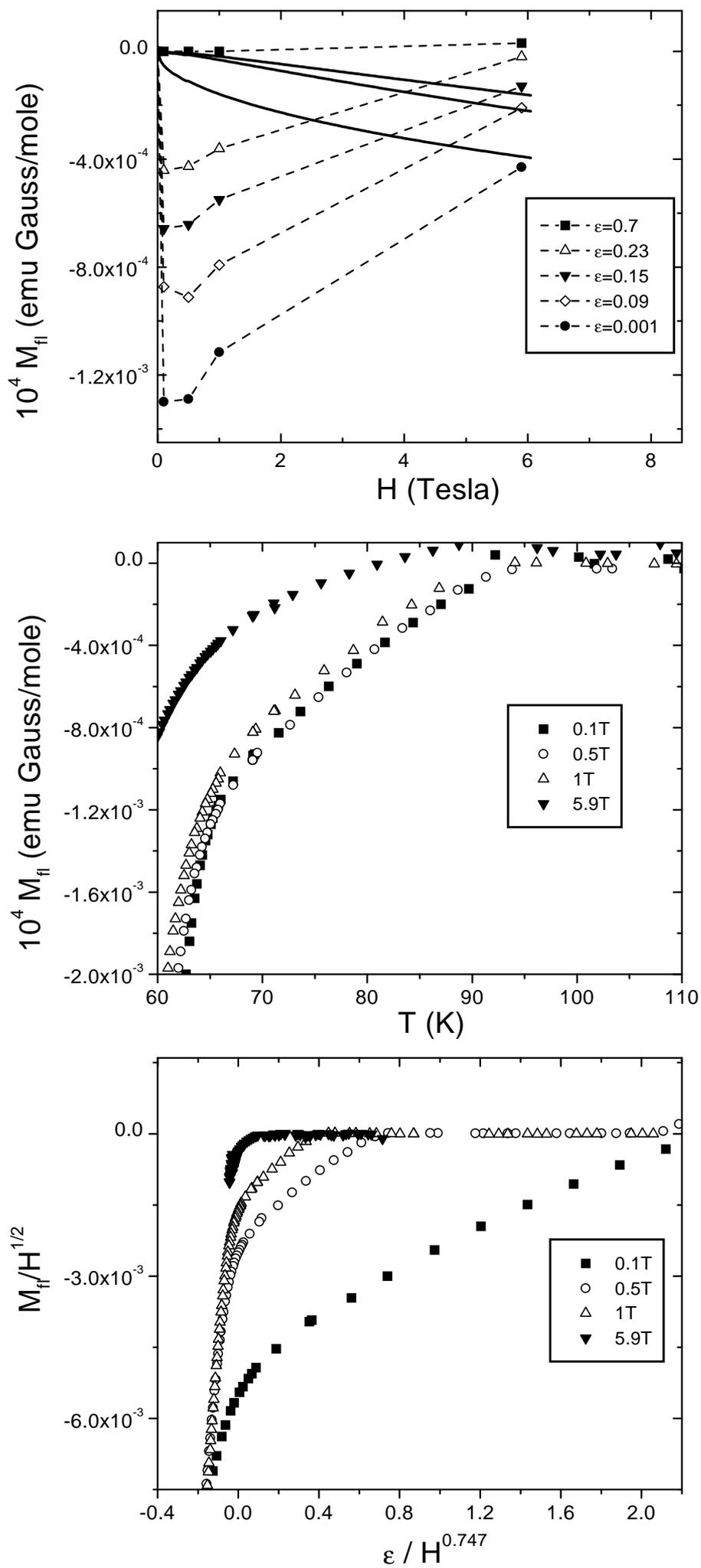
\begin{figure}
\caption {  Evidence of the failure of anisotropic GL functional and of 
scaling arguments to describe the magnetization curves in underdoped YBCO.
The solid lines in the upper part of the figure are the behaviours
obtained from the numerical derivative of Eq. 2.
 The dashed lines are guides for the eye.}
\end{figure}
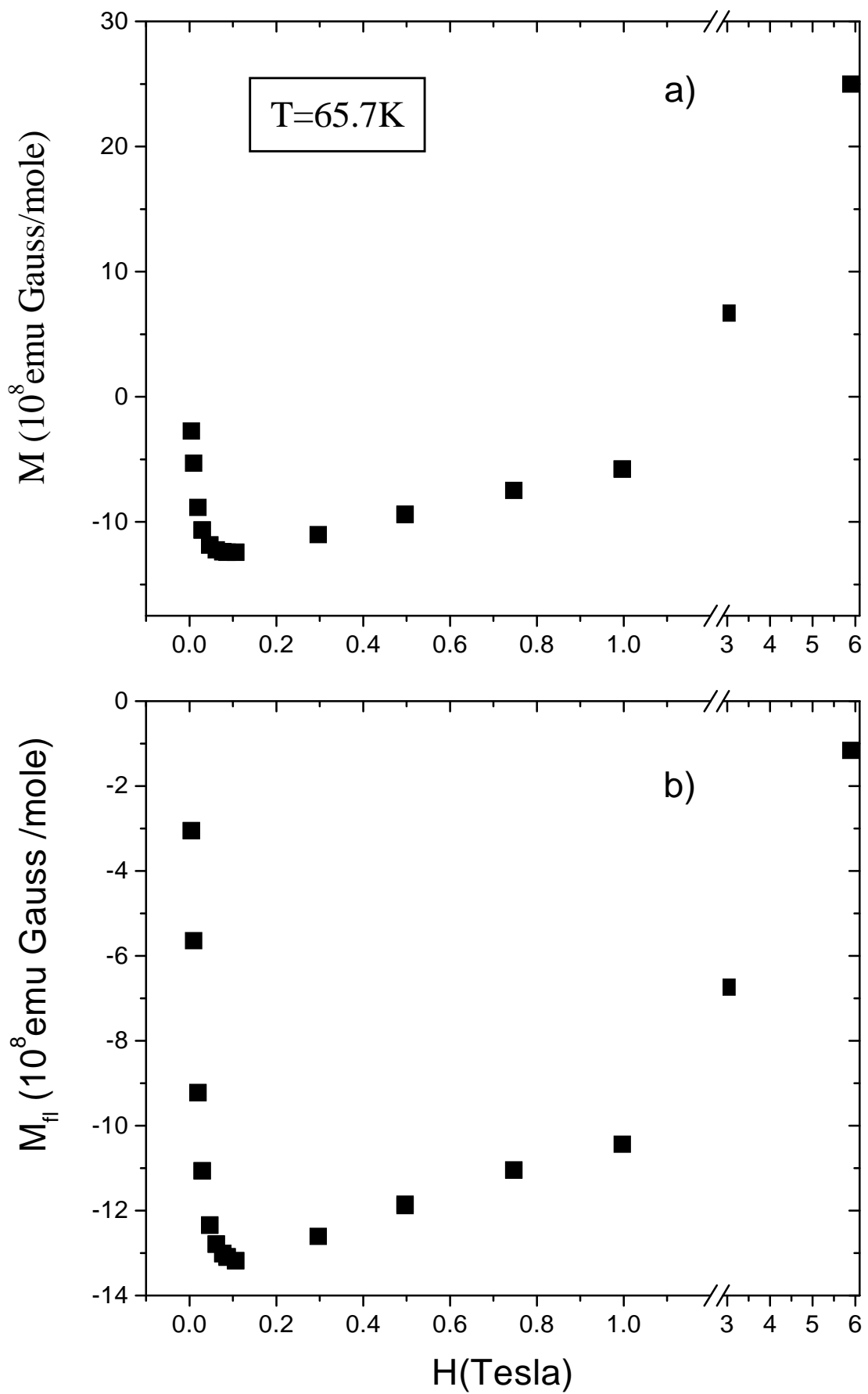
\begin{figure}
\caption { Magnetization curve in chain-ordered underdoped YBCO 
under high stabilization temperature ( $T= 65.67 K$ ). {\it a)}
 raw data for the total magnetization, {\it b)} 
fluctuating magnetization $M_{fl}$ as a  function of the field.}
\end{figure} 
\newpage
 
 
{\small
\begin{center}
\begin{tabular}{|c|c|c|c|c|}
\hline
 &YBCO  &  \multicolumn{3}{|c|}{YBCO}  \\
 &optimally doped  &  \multicolumn{3}{|c|}{underdoped}  \\
 & &  disordered & \multicolumn{2}{|c|}{chain-ordered}   \\
 \hline
   $T_c(0)$ & $91.1 \pm 0.4 \; K$   & $63.3 \pm 0.4 \;
K$ & $64.5 \pm 0.4 \; K$ 
  &  $64.2 \pm 0.4 \; K$  \\
 \hline
 $T_c(H= 6 Tesla)$ & $87 \pm 0.5 \; K$  & $60 \pm 1 \; K $ &
    -  &  - \\
 \hline
 $T_c(H= 9.4 Tesla)$ & -   & $55 \pm 1 \; K $ &
    $54 \pm 1 \; K $   &  - \\
 \hline
  $\left( \frac{\partial H_{c2}}{\partial T} \right)_{T_{c0}}$ $T/K$
  & $-1.2 \pm 0.2  $   & $ -1.05 $ &
  $ -1.1 \pm 0.2  $ & $-1.0 \pm 0.1  $\\
 \hline
 $H_{c2}(T \rightarrow 0) \quad  \bf{H} \parallel \bf {c} $ 
 & $100 \pm 20 \; T$  &  $\sim 70 \; T $ &
  $\sim 65 \;T $ & - \\
 \hline
  $H_{c2}(T \rightarrow 0) \quad  \bf{H} \perp \bf {c} $ 
 & $ \sim 67  \; T $   & - &
  $ \sim 44 \; T $ & - \\
\hline
\end{tabular} 
\end{center}
}




\begin{references}
 
 
\bibitem{bib1} For a recent collection of papers,  see '' Fluctuations
Phenomena in High    
Temperature Superconductors'',  Edited  by  M. Ausloos and A.A. Varlamov,
Kluwer Academic Publisher,   
(1997)
\bibitem{bib2} A.A. Varlamov,  G.Balestrino,  E. Milani and D.V. Livanov,
Advances in Physics 48, 655 (1999) 
\bibitem{bib3b} M.Thinkham ``Introduction to
Superconductivity'', Mc Graw-Hill N.Y. (1996) Chapters 8-9  
\bibitem{bib3} See A.Junod,  in ``Studies of High Temperature
Superconductors'',  edited by A.V. 
Narlikar ( Nova Science Publishers,  Inc. New York, 1996 ), Vol. 18
\bibitem{bib4} V.Pasler et al. Phys. Rev. Lett.  81, 1094 (1998)
\bibitem{bib5} Z.H. Lin et al. Europhys. Letters 32, 573 (1995) and
references therein
\bibitem{bib6} M.A. Howson et al. Phys. Rev. B 51, 11984 (1995) and
references therein
\bibitem{bib7} See the review paper by L.N.Bulaevskii, Int. J. of Modern
Physics B4, 1849 (1990) and  several papers in Ref. 1 
\bibitem{bib8} P.Carretta, D. Livanov, A.Rigamonti and A.A.Varlamov,
Phys.Rev.B 54,  R9682 (1996)    
\bibitem{bib9} P.Carretta, A.Rigamonti, A.A.Varlamov, D.Livanov, Nuovo
Cimento 19 , 1131 (1997) 
\bibitem{bib10} V. Mitrovic et al., Phys. Rev. Lett.   82, 2784 (1999)
\bibitem{bib11} H.N.Bachman et al., Phys. Rev. B 60, 7591 (1999)
\bibitem{bib12} K. Gorny et al., Phys. Rev. Lett. 82, 177 (1999)
\bibitem{bib15} M. Eschrig et al.,  Phys. Rev. B 59, 12095 (1999)
\bibitem{bib16}  M.Randeira and A.A.Varlamov,  Phys. Rev. B 50,  10401 (1994)
\bibitem{bib17} A.Rigamonti, F.Borsa and P.Carretta, Rep.Prog.Phys.61, 1367
(1998)
\bibitem{bib18} A.G. Loeser et al. Science 273, 325 (1996)
\bibitem{bib19} H.Ding et al. Nature 382,  51 ( 1996)
\bibitem{bib19b} G. V. M. Williams, J. L. Tallon, R. Michalak and R. Dupree, Phys. Rev. B 54,
R6909 (1996)
\bibitem{bib20} T. Dahm,  D. Manske and L. Tewordt,  Phys. Rev. B 55,  15274
(1997)
\bibitem{bib21} V.J.Emery, S.A.Kivelson and O.Zachar Phys.Rev. B 56, 6120
(1997); see also V. J. Emery and S. A. Kivelson, J. Phys. Chem. Sol. 59,
1705 (1998) 
\bibitem{bib22} C. Castellani,  C. Di Castro and M. Grilli  Z.Phys. B 103,
137 (1997)
\bibitem{bib23} C.Baraduc,  A. Budzin,  J-Y Henry,  J.P. Brison and L.
Puech,  Physica C 248
	138 (1995 )
\bibitem{bib24} A.Budzin and V.Dorin, Ref. 1, p.335
\bibitem{bib25} M. B. Salamon et al. Phys.Rev. B 47,  5520 (1993)
\bibitem{bib26} T.Schneider and H. Keller,  Inter. Jornal of Modern Physics
B 8,  487 (1993)
\bibitem{bib27} A.Junod,  J.Y. Genoud,  G. Triscone and T. Schneider,
Physica C 294, 115 (1998)
\bibitem{bib28} R.E. Prange,  Phys. Rev. B 1,  2349 (1970) 
\bibitem{bib29}  M.A.Hubbard, M.B. Salamon and B.W.Veal, Physica C 259, 309
(1996) 
\bibitem{bib30} C.P.Slichter ``Principles of Magnetic Resonance'', Springer
Verlag, Berlin-Heidelberg 
   (1990)
\bibitem{bib31} P.Manca, P.Sirigu, G.Castellani and A.Migliori, Il Nuovo
Cimento 19, 1009 (1997); G.Castellani, A.Migliori, P.Manca and P.Sirigu, Il Nuovo
Cimento 19, 1075 (1997)
\bibitem{bib36} U.Welp et al., Phys.Rev.Lett. 67, 3180 (1991)
\bibitem{bib39} A.I.Buzdin and A.A.Varlamov,  Phys.Rev.B 58, 14195 (1998) 
\bibitem{bib40} P.Mosconi, A.Rigamonti and A. A. Varlamov, Appl. Mag. Reson. to be published
\bibitem{bib36b} Yu. N. Ovchinikov, S. A. Wolf and V. Z. Kresin, Phys. Rev. B 60, 4329 (1999)
\bibitem{bib37b} E. Z. Kuchinskii and M. V. Sadovskii, cond-mat/9910261
\bibitem{bib45} D.Mihailovic, T.Mertelj and K.A.Muller, Phys. Rev. B  57,
6116 (1998); 
 see also Kabanov et al. Phys.Rev.B  59, 1497 (1999) 
\bibitem{bib48} M. R. Norman, cond-mat/9904048
\bibitem{bib49} A.Bianconi et al. Phys.Rev.Lett. 76, 3412 (1996)
\bibitem{bib41b} C. Bergemann et al., Phys. Rev. B 57, 14387 (1998)


 
 






\end{references}
\end{document}